\documentclass{article}
\usepackage{spconf,amsmath,graphicx,hyperref}
\usepackage{cite}
\usepackage{amsmath,amssymb,amsfonts}
\usepackage{algorithmic}
\usepackage[ruled,vlined]{algorithm2e}
\usepackage{graphicx}
\usepackage{textcomp}
\usepackage{xcolor}
\usepackage{amsmath,graphicx}
\usepackage{pifont}
\usepackage{hyperref}       
\usepackage{url}            
\usepackage{booktabs}       
\usepackage{amsfonts}       
\usepackage{nicefrac}       
\usepackage{microtype}      
\usepackage{lipsum}
\usepackage{graphicx}
\usepackage{amsmath}
\usepackage{caption}
\usepackage{subcaption}
\usepackage{multirow}
\graphicspath{ {./images/} }
\usepackage{amsfonts}       
\usepackage{color, colortbl, arydshln}
 
\newcommand{\myparagraph}[1]{\noindent\textbf{#1} \hspace{.15cm}} 
\def\BibTeX{{\rm B\kern-.05em{\sc i\kern-.025em b}\kern-.08em
    T\kern-.1667em\lower.7ex\hbox{E}\kern-.125emX}}

\renewcommand{\baselinestretch}{0.91} 

\title{Residual Tokens Enhance Masked Autoencoders for Speech Modeling}
\name{Samir Sadok$^{\star}$ \qquad Stéphane Lathuilière$^{\star}$ \qquad Xavier Alameda-Pineda$^{\star}$}  
\address{$^{\star}$Inria at Univ. Grenoble Alpes, CNRS, LJK, France}

\begin{document}
\ninept
\maketitle
\begin{abstract}
Recent speech modeling relies on explicit attributes such as pitch, content, and speaker identity, but these alone cannot capture the full richness of natural speech. We introduce RT-MAE, a novel  masked autoencoder framework that augments the supervised attributes-based modeling with unsupervised residual trainable tokens, designed to encode the information not explained by explicit  labeled factors (e.g., timbre variations, noise, emotion etc). Experiments show that RT-MAE improves reconstruction quality, preserving content and speaker similarity while enhancing expressivity. We further demonstrate its applicability to speech enhancement, removing noise at inference while maintaining controllability and naturalness\footnote{Code and audio examples are available online: \url{https://samsad35.github.io/site-residual}}.
\end{abstract}
\section{Introduction}
\label{sec:intro}

Extensive research has focused on modeling speech signals, ranging from parametric models such as linear predictive coding (LPC) \cite{markel1976linear}, sinusoidal models \cite{mcaulay1986speech, george1997speech}, and harmonic-pulse-noise models \cite{moulines1990pitch}
 and the phase vocoder \cite{laroche1993hns}. Vocoders such as STRAIGHT \cite{kawahara2006straight} and WORLD \cite{morise2016world} further enabled real-time speech manipulation by decomposing signals into interpretable acoustic features (e.g., $f_0$, spectral envelope, aperiodicity). More recently, deep learning has become the dominant paradigm, with neural audio codecs compressing speech into low-bitrate discrete units \cite{zeghidour2021soundstream, defossez2023high}, self-supervised encoders such as HuBERT \cite{hsu2021hubert} capturing linguistic content for reconstruction when combined with pitch and speaker identity \cite{polyak2021speech} achieving high-quality synthesis, or AnCoGen \cite{sadok2025ancogen} that uses the masked modeling mechanism to learn a bidirectional mapping between Mel-Spectrogram tokens and high-level attributes (e.g., pitch, content, speaker identity, loudness, etc), enabling both analysis (attributes from Mel-Spectrograms) and generation (Mel-Spectrograms from attributes).
 
However, the speech signal is more complex than what explicit attributes can represent. Factorizing only content, speaker identity, and prosody misses much of the richness of natural speech. The \emph{residual}—defined here as the aspects of speech that cannot be described by the available explicit attributes—encompasses speaking style, emotion, articulatory nuances, micro-prosody, and non-verbal cues, yet it remains largely unmodeled. Most recent methods \cite{sadok2025ancogen, polyak2021speech, nguyen2025spirit, tan2025seeing} rely solely on clearly defined attributes (e.g., $f_0$, linguistic features, speaker embeddings), while residual information is implicitly absorbed as a dataset-specific bias, limiting generalization and control. 

Early source-filter models \cite{markel1976linear, kawahara2006straight, morise2016world} already pointed out the presence of a \emph{residual} component, corresponding to the excitation signal. This residual carries irregularities such as jitter, shimmer, or aperiodicity that cannot be fully explained by the pitch or timbre. In addition, it also conveys subtle expressive information that enriches speech beyond explicit attributes. Later research investigated this residual more systematically: disentanglement-based models separate content, prosody, and speaker identity, while the remaining variability is interpreted as residual factors. Examples include factorized VAEs (FHVAE~\cite{hsu2017unsupervised}), adversarially regularized models~\cite{hsu2019disentangling}, hierarchical VAE-based prosody models~\cite{hsu2018hierarchical}, Global Style Tokens (GST)\cite{wang2018style}, and voice conversion frameworks like AutoVC~\cite{qian2019autovc} and StarGAN-VC2~\cite{kaneko2019stargan}. These latent representations capture rhythm, energy, emotion, and other expressive cues, complementing explicit attributes to improve naturalness. However, most of these approaches rely on complex disentanglement objectives and multiple task-specific losses, which make the resulting models computationally demanding, less generalizable, and often restricted to narrow applications such as voice conversion or style transfer.

To address these limitations, we propose RT-MAE, a residual-token masked autoencoder for speech modeling that explicitly captures residual information while leveraging recent advances in speech representation learning \cite{polyak2021speech, sadok2025ancogen, nguyen2025spirit, tan2025seeing}. In RT-MAE, residual information is represented as continuous tokens directly integrated in an MAE \cite{he2022masked}, following the design of~\cite{sadok2025ancogen}, which supports speech analysis, control, and generation.  Unlike prior disentanglement-based approaches, this design provides a flexible representation of residual information. 
Our main contributions are:
 (1) we introduce residual tokens to explicitly encode the \emph{rest} of speech information not captured by attributes (e.g., timbre variations, emotion, micro-prosody);
 (2) we propose a regularization mechanism to control the information flow in these tokens, ensuring a balanced trade-off between reconstruction quality and attributes controllability;
 and (3) we show that these residual tokens enable various manipulation tasks, and in this work, we focus on noise modeling, where noise can be captured and then deactivated at inference to achieve speech denoising.

\begin{figure*}[!ht]
    \centering
    \includegraphics[width=0.87\linewidth]{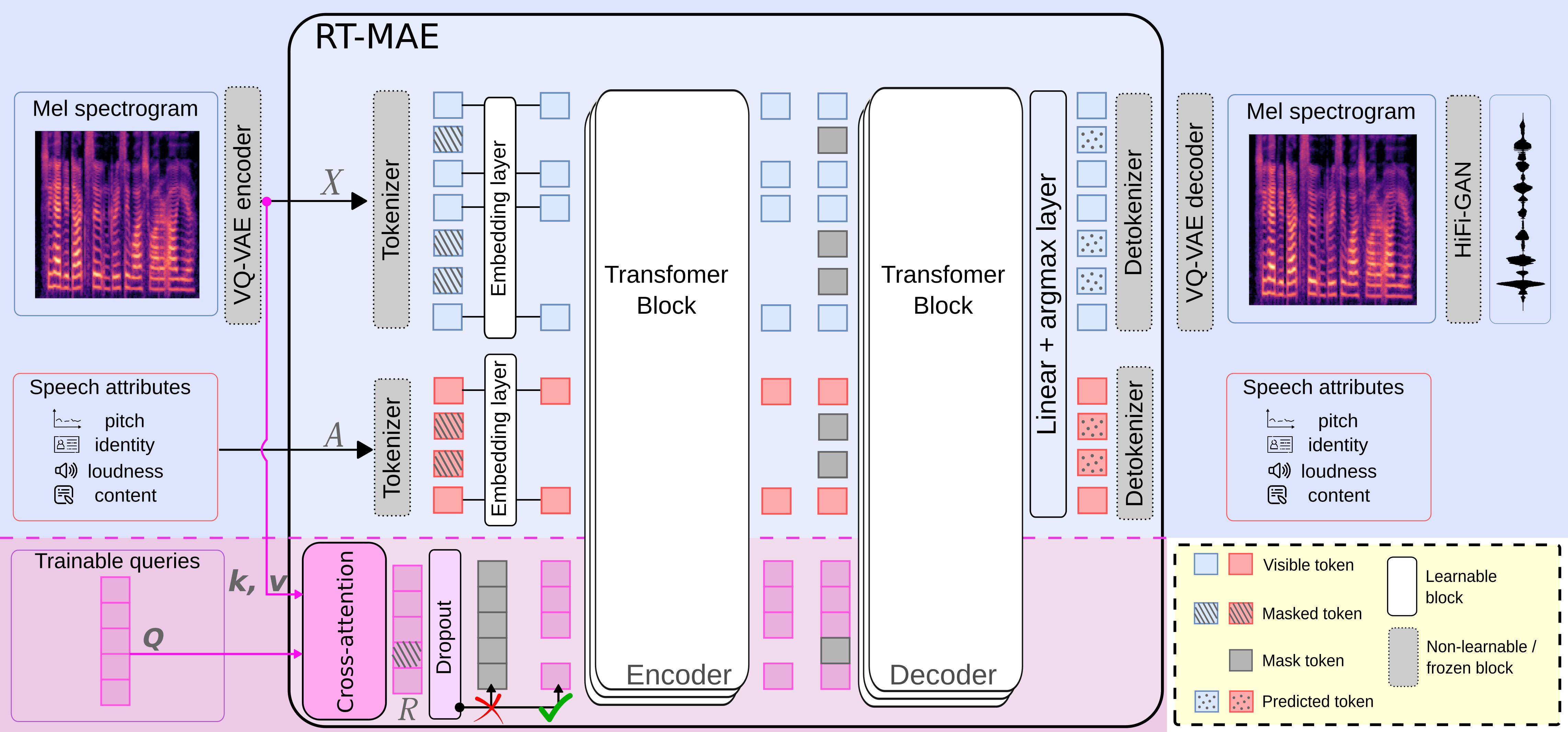}
    \caption{Overview of the proposed RT-MAE architecture. The top (blue) illustrates the current paradigm, where speech and explicit attributes are jointly modeled using an MAE. The bottom (purple) illustrates the introduction of trainable queries, which through cross-attention, capture residual factors not captured by the explicit attributes.
}
    \vspace{-6pt}
    \label{fig:AnCoGen_archi}
    \end{figure*}

\section{The proposed approach: RT-MAE}

In this section, we present the methodology of our approach.
We begin by introducing the notation and providing intuition on why explicit attributes alone are insufficient to fully model speech (\ref{sec:notation_intuition}).
We then describe how we construct residual tokens using a cross-attention mechanism inspired by the Perceiver architecture (\ref{subsec:perceiver}), allowing us to encode complementary information in a compact and trainable form.
Finally, we explain how we control the flow of information in these residual tokens through a dropout-based regularization strategy (\ref{subsec:information}), which prevents redundancy, encourages the effective use of explicit attributes, and ensures that speech generation remains interpretable and controllable.

\subsection{Background and Intuition}
\label{sec:notation_intuition}
Let $\mathbf{X} \in \mathbb{R}^{T \times F}$ denote the Mel-Spectrogram of a speech utterance with $T$ frames and $F$ frequency bins, and let $\mathbf{A} = {A_1, \dots, A_K}$ represent explicit speech attributes such as pitch, linguistic content, speaker embedding, or loudness. Generating $\mathbf{X}$ from $\mathbf{A}$ alone provides interpretable control but cannot capture all aspects of natural speech: with a finite set of attributes, many subtle phenomena remain unmodeled. Fully explaining speech would require an impractically large (or even unbounded) set of attributes. To address this, we introduce $N$ continuous residual tokens $\mathbf{R} \in \mathbb{R}^{N \times d}$ that encode information not captured by $\mathbf{A}$, where $d$ denotes their dimensionality. The model reconstructs speech from both $\mathbf{A}$ and $\mathbf{R}$, separating structured control from residual variability and enabling both controllability and flexibility beyond attribute-only models. 
How can these tokens $\mathbf{R}$ be estimated to  effectively complement the explicit attributes?

Figure~\ref{fig:AnCoGen_archi} illustrates the overall RT-MAE architecture, which builds upon the AnCoGen framework~\cite{sadok2025ancogen} using an MAE architecture. In our approach, the Mel-Spectrogram $\mathbf{X}$ and attributes $\mathbf{A}$ are quantized into discrete tokens, while the residual tokens $\mathbf{R}$ remain continuous.  Following the MAE framework, all three modalities are partially masked during training, encouraging the model to learn intra- and inter-dependencies. Visible discrete tokens are embedded into continuous vectors via trainable codebooks. The three types of embeddings are concatenated into a single sequence, which is then processed by the Transformer encoder. After encoding, mask tokens are inserted to complete the sequence before decoding. At inference, masking is task-dependent: for analysis, attributes are predicted from an input Mel-Spectrogram, while for generation, the Mel-Spectrogram is reconstructed from attributes and the residual tokens, with the final waveform synthesized using a HiFi-GAN vocoder~\cite{kong2020hifi}. The autoencoder is trained by minimizing the cross-entropy loss between predicted and ground-truth tokens.

A central question is how to integrate the residual $\mathbf{R}$ in the MAE architecture, addressing both its architectural role in complementing discrete tokens (Section~\ref{subsec:perceiver}) and the training strategy to regulate information flow (Section~\ref{subsec:information}).

\subsection{Extracting the residual tokens via cross-attention}
\label{subsec:perceiver} 

Our goal is to learn residual tokens that encode the residual information of the Mel-Spectrogram. A natural choice for this is an attention-based mechanism, which can selectively aggregate information across time and frequency. However, standard self-attention requires a token for every Mel frame, which significantly increases computational cost. Moreover, we want to control the number of trainable tokens that encode the residual, effectively acting as a compression bottleneck.

Inspired by the Perceiver architecture \cite{jaegle2021perceiver}, we address these challenges by introducing a fixed-size set of learnable queries $\mathbf{Q}$ that interact with the input through \emph{cross-attention}. Each vector queries the entire Mel-Spectrogram, attends to the most relevant information, and aggregates it into a compact representation. This mechanism allows the model to selectively encode residual factors without needing a one-to-one mapping between residual and input tokens. After cross-attention, the token vectors are further refined through the Transformer encoder as represented in Figure~\ref{fig:AnCoGen_archi}.

Formally, the learnable queries $\mathbf{Q}$ attend to the input embedding $\mathbf{X}$, which provides keys ($\mathbf{K}$) and values ($\mathbf{V}$). Attention weights are computed by measuring the similarity (typically via dot product) between each query and all input keys: $\mathbf{M} \gets \text{softmax}\!\left( \frac{\mathbf{Q}\mathbf{K}^\top}{\sqrt{d}} \right)$, then used to form a weighted sum over values, producing the residual tokens $\mathbf{R} \gets \mathbf{M}^T\mathbf{V}$. Each token can thus focus on different aspects of the input, compressing the Mel-Spectrogram into a compact space.  

Incorporating these residual tokens into the MAE paradigm \cite{sadok2025ancogen} enables the model to complement explicit attributes with residual information (as shown in Figure~\ref{fig:AnCoGen_archi}). The residual space thus acts as a bottleneck, organizing the remaining degrees of freedom in the speech signal and facilitating the generation of richer, more natural, and controllable speech outputs.

\subsection{Regularization strategy}
\label{subsec:information}

\begin{figure*}[t]
\centering
    \centering
    \resizebox{.8\linewidth}{!}{ 
    \begin{tabular}{ccccccccccc}
    \toprule
         & &\multicolumn{4}{c}{\textbf{LibriSpeech Test} \cite{panayotov2015librispeech}} & \multicolumn{4}{c}{\textbf{EmoV-DB} \cite{adigwe2018emotional}} \\
        \cmidrule(lr){3-6} \cmidrule(lr){7-11}
         & \# Parameters (M) &STOI \footnotesize{$\uparrow$} & N-MOS \footnotesize{$\uparrow$} & SBS \footnotesize{$\uparrow$} & COS \footnotesize{$\uparrow$} & STOI \footnotesize{$\uparrow$} & N-MOS \footnotesize{$\uparrow$} & Acc. \footnotesize{$\uparrow$} & COS \footnotesize{$\uparrow$}& ~ \\ \midrule
        \color{gray} GT MS & \color{gray} - & \color{gray} 0.93 & \color{gray} 4.44 & \color{gray} - & \color{gray} 0.96 & \color{gray} 0.93 & \color{gray} 4.40 & \color{gray} 99.30 & \color{gray} 0.94 & ~ \\ 
        AnCoGen~\cite{sadok2025ancogen} & 27.7 & 0.77 & 4.04 & 0.83 & 0.81 & 0.70 & 4.23 & 96.79 & 0.80 & ~ \\
        \rowcolor{blue!10}
        RT-MAE (Ours) & 28.9 & \textbf{0.82} & \textbf{4.32}  & \textbf{0.86} & \textbf{0.92} & \textbf{0.76} & \textbf{4.31} & \textbf{98.65} & \textbf{0.88} & ~ \\
        \bottomrule
    \end{tabular}
    }
    \captionof{table}{\textbf{Speech analysis and synthesis results} on LibriSpeech and EmoV-DB test sets. GT MS refers to the ground-truth Mel-Spectrogram, which is converted to waveform using the HiFi-GAN vocoder (without passing through the model).}
    \vspace{-6pt}
    \label{tab:synthesis}
\end{figure*}
 
Without regularization, the residual tokens could encode all the information needed to reconstruct the Mel-Spectrogram—essentially reducing the model to a simple encoder-decoder with very limited structuring and generalization capabilities. 
In this case, the model might rely primarily on the residual tokens, which do not use the attributes, limiting controllability and compromising interpretability in speech generation.

To address this issue, we introduce a dropout-based regularization mechanism applied to the residual tokens. During training, we draw a value uniformly from $[0,1]$ and compare it to a fixed threshold $\tau \in [0,1]$. If the sampled value is below $\tau$, the residual tokens are dropped, i.e., they are not provided to the transformer, forcing the model to generate the Mel-Spectrogram solely from the explicit attributes. Otherwise, the residual tokens are concatenated with the attributes. 
Note that, differently from regular dropout~\cite{srivastava2014dropout} the entire vector is dropped and not only single neurons. 
This strategy prevents the model from over-relying on the residual tokens, encourages effective utilization of the provided attributes, and ensures that the generated speech remains both interpretable and controllable. 


\section{Experiments}

\subsection{Experimental setup}
\label{subsec-exp-setup}

\myparagraph{Settings} To ensure a fair comparison, we follow the training protocol of AnCoGen and train our model using the LibriSpeech 360 Clean dataset \cite{panayotov2015librispeech}. The training data was preprocessed to extract four types of attributes: Pitch (using CREPE \cite{kim2018crepe}), Loudness using the root-mean-square energy (RMSE), speaker identity embeddings obtained from a pretrained speaker encoder \cite{desplanques2020ecapa}, and     content represented by phonetic posteriorgrams (PPGs \cite{churchwell2024high}) derived from a forced-alignment model. The model was trained using 4 NVIDIA A100 GPUs, with a batch size of 128 and the AdamW optimizer, for a total of 400 epochs.\\

\myparagraph{Architecture} The architecture of RT-MAE used in this work builds upon the original AnCoGen model \cite{sadok2025ancogen}. Both the encoder and decoder consist of 6 Transformer layers, each comprising multi-head self-attention mechanisms, position-wise feedforward networks, and layer normalization. We introduce $N=25$ trainable vectors, each having the same dimensionality as the internal representations used in the Transformer ($d=512$). These vectors interact with the Mel-Spectrogram via cross-attention (see Section~\ref{subsec:perceiver}), allowing the model to capture complementary information beyond the conditioning attributes. Unless otherwise specified, the dropout threshold controlling the use of residual tokens are fixed at $\tau=0.5$.\\

\myparagraph{Metrics}
To evaluate the quality and intelligibility of the generated speech, we rely on a combination of objective and perceptual metrics. STOI (Short-Time Objective Intelligibility) \cite{taal2010short} measures speech intelligibility, with higher values indicating better performance. N-MOS (Neural Mean Opinion Score) non-intrusive estimate of speech naturalness (1–5) using pretrained models such as SQUIM \cite{kumar2023torchaudio}. speechBertScore (SBS) \cite{saeki2024speechbertscore} is computed by passing the generated speech through an ASR system, with higher scores indicating better intelligibility and linguistic accuracy. Finally, COS (Cosine Similarity on speaker embeddings) \cite{wan2018generalized} measures the similarity between speaker embeddings of the generated and reference speech, providing an estimate of speaker consistency.

\subsection{Results and discussion}

\myparagraph{Improving speech synthesis through the residual tokens}
Table~\ref{tab:synthesis} presents the synthesis results, comparing the original model relying solely on explicit attributes~\cite{sadok2025ancogen} with our version augmented by the residual tokens. In this experiment, each speech signal is first analyzed to extract its attributes, and then reconstructed either using only these attributes (as in~\cite{sadok2025ancogen}) or with the additional information provided by the residual tokens.
The results in Table~\ref{tab:synthesis} highlight the benefits of introducing the residual tokens in our framework. On the Librispeech test set, RT-MAE consistently improves AnCoGen.  STOI increases by +0.05, indicating better intelligibility, while N-MOS also improves by +0.28, showing gains in perceptual quality. Furthermore, SpeechBertScore rises by +0.03, and speaker similarity achieves a notable gain of +0.09, confirming a more faithful preservation of linguistic and speaker information.
On Emov-DV \cite{adigwe2018emotional}, which is more challenging due to its expressive and emotional content, RT-MAE also outperforms AnCoGen across all metrics. STOI improves from 0.7 to 0.76, and N-MOS increases from 4.18 to 4.31, reflecting both high intelligibility and quality emotional speech. More importantly, emotion classification accuracy (96.79$\rightarrow$98.65) and cosine similarity (0.70$\rightarrow$0.88) show that our model preserves emotional cues and speaker identity more effectively.
This raises a key question: \textit{what do the residual tokens actually encode? Does it capture complementary information beyond the explicit attributes, or is it redundant with them and potentially conflicting?} We address this question in the following.\\

\begin{table}[t]
    \centering
    \caption{\textbf{Speech analysis and synthesis (ablation study):} Uncovering the information encoded in the residual tokens. Results are reported at inference. A cross (\ding{55}) indicates that the component is masked, whereas a check (\ding{51}) indicates it is not masked.}
    \label{tbl-Information}
    \resizebox{1.\linewidth}{!}{ 
    \begin{tabular}{c c c c c c}
    \toprule
    Attributes ($\mathbf{A}$) & Residual tokens ($\mathbf{R}$) & STOI $\uparrow$ & N-MOS $\uparrow$ & SBS $\uparrow$ & COS $\uparrow$ \\
    \midrule
    \ding{55} & \ding{55} & 0.27 & 2.32 & 0.44 & 0.50 \\
    \ding{51} & \ding{55} & 0.76 & 4.03 & 0.83 & 0.81 \\
    \ding{55} & \ding{51} & 0.50 & 3.04 & 0.56 & 0.72 \\
    \ding{51} & \ding{51} & \textbf{0.82} & \textbf{4.32} & \textbf{0.86} & \textbf{0.92} \\
    \bottomrule
    \end{tabular}
    }
    \vspace{-6pt}
\end{table}

\myparagraph{Analyzing the complementarity between residual tokens and attributes} 
Since RT-MAE uses a Transformer-based masked modeling approach to predict masked tokens, we can vary the masking rate to selectively disable explicit attributes and/or the residual tokens. This setup enables us to investigate the role of the residual tokens in speech synthesis, specifically how it complements or compensates for missing attribute information. Table~\ref{tbl-Information} presents the synthesis results under different masking configurations, where attributes and/or the residual tokens are entirely removed.
When audio synthesis is performed without the additional tokens at inference (second row), the architecture reduces to AnCoGen (second row in Table \ref{tab:synthesis}), differing only in the training procedure, as our model is trained with the residual tokens. The table shows that the results closely match those of AnCoGen \cite{sadok2025ancogen}, confirming that our model effectively leverages the attributes for synthesis.
In contrast, the third row (where attributes are removed and only the residual tokens are used) shows a \textit{substantial} drop in both content and quality, as expected. Nevertheless, the residual tokens still achieves a relatively high speaker similarity score, suggesting that it encodes meaningful identity information.
Finally, the last row, which combines both attributes and the residual tokens during synthesis (RT-MAE), leads to consistent improvements across all similarity metrics. This highlights the role of the $\mathbf{R}$ as a residual component that complements the attributes and further enhances synthesis quality.\\

\begin{figure}[t!]
    \centering
    \includegraphics[width=\linewidth]{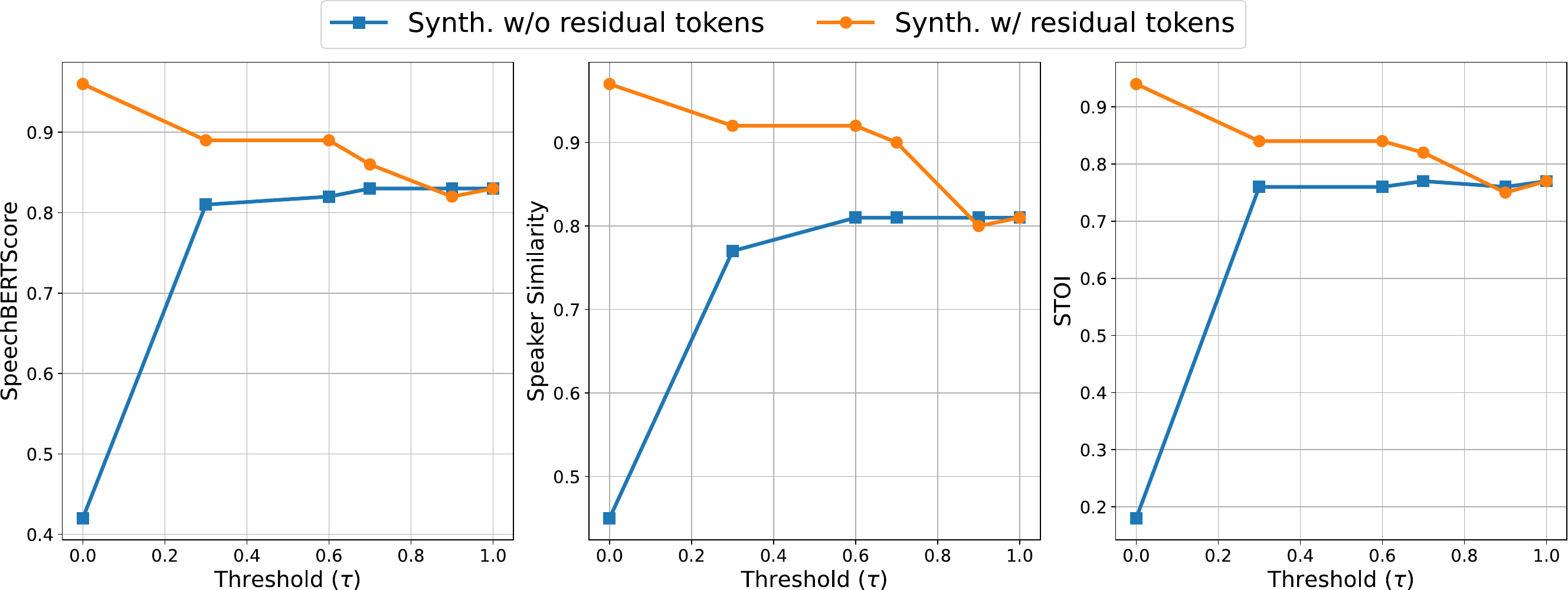}
    \caption{Effect of the dropout threshold $\tau$ on synthesis quality. 
    In blue (\textcolor{blue}{\rule[0.5ex]{1em}{1.pt}}), only the attributes are used ($\mathbf{R}$ is masked). In orange (\textcolor{orange}{\rule[0.5ex]{1em}{1.pt}}), both attributes and residual tokens are used.
    }
    \vspace{-6pt}
    \label{fig:drop-out}
\end{figure}

\myparagraph{Adjusting the information flow through residual tokens} 
This paragraph examines how the dropout threshold ($\tau$) shapes the balance between attributes and the residual tokens. We train models with $\tau$ ranging from 0 (residual tokens always available) to 1 (residual tokens completely disabled). As shown in Figure~\ref{fig:drop-out}, with $\tau=0$ the model relies almost entirely on the residual tokens, ignoring attributes and losing controllability.
As $\tau$ increases (starting around 0.5), the model begins to effectively integrate both the attributes and the residual tokens, resulting in improved content preservation, speaker similarity, and overall synthesis quality.
For high values ($\tau > 0.8$), the residual tokens are largely disregarded, and synthesis depends almost solely on attributes.\\

\myparagraph{Preserved Controllability} In this paragraph, we report pitch manipulation experiments showing that the introduction of residual tokens preserves the controllability capacity of AnCoGen. Table~\ref{tab:pitch-manipulation} reports pitch manipulation results on the PTDB dataset~\cite{pirker2012pitch} with ground-truth pitch values. We compare our method without and with the residual tokens for target shifts of 0\%, 10\%, and 20\%. Evaluation is based on Absolute Average Error (AAE, lower is better), naturalness (N-MOS), and SpeechBertScore (SBS). Results show that AAE remains essentially unchanged, confirming that pitch control accuracy is preserved,  while the inclusion of the residual tokens consistently improves N-MOS and SBS across all target shifts. This demonstrates that adding the residual token enhances perceptual quality and semantic consistency, while maintaining precise and controllable pitch manipulation.\\
 
\myparagraph{Encoding noise through an additional residual token} Previously, we showed that the residue captures information beyond explicit attributes.
We now extend this idea where the noise is treated as a specific form of residual information. To this end, we introduce an additional residual vector, $\mathbf{R}_{\text{noise}}$, dedicated to modeling noise. Assuming annotations indicating the presence of noise, this residue is activated only when the input contains noise, while the previous residue $\mathbf{R}$ continues to capture residual speech-related information. To enforce disentanglement \cite{wang21n_interspeech}, we minimize the mutual information between the two residues ($\mathbf{R}$ and $\mathbf{R}_{\text{noise}}$) using the CLUB estimator \cite{cheng2020club}, ensuring that each encodes complementary information.
For evaluation, we use LibriMix \cite{cosentino2020librimix}, which combines clean LibriSpeech utterances \cite{panayotov2015librispeech} with WHAM! noise signals \cite{wichern2019wham}. At inference time for noise reduction, speech attributes are extracted as usual, but the $\mathbf{R}_{\text{noise}}$ is deactivated during synthesis, effectively removing noise while preserving the residual speech characteristics via $\mathbf{R}$. Additionally, we report DNSMOS \cite{reddy2022dnsmos}, which provides objective estimates of the speech signal quality (SIG), background noise intrusiveness (BAK), and overall quality (OVRL).
Table~\ref{tab:noise} reports speech denoising results on the LibriMix test set. RT-MAE consistently improves objective metrics compared to the original AnCoGen. In particular, it achieves the highest N-MOS (4.25) and SIG (4.23) scores, while maintaining strong BAK (4.29) and OVRL (3.80) performance, confirming improved perceptual quality and robustness to noise. The COS score increases from 0.73 to 0.86, significantly narrowing the gap with task-specific models like Conv-TasNet.

\begin{table}[t]
\captionof{table}{\textbf{Pitch manipulation results} on the PTDB dataset~\cite{pirker2012pitch}.}
\label{tab:pitch-manipulation}
\centering
\resizebox{1.0\linewidth}{!}{ 
\begin{tabular}{ccccccccccc}
\toprule
\multicolumn{2}{c}{} & \multicolumn{3}{c}{+0 \%} & \multicolumn{3}{c}{+10 \%} & \multicolumn{3}{c}{+20 \%} \\ 
\cmidrule(lr){3-5} \cmidrule(lr){6-8} \cmidrule(lr){9-11}
 \multicolumn{2}{c}{$\mathbf{R}$}  & AAE \footnotesize{$\downarrow$} & N-MOS \footnotesize{$\uparrow$}  & SBS \footnotesize{$\uparrow$}  & AAE \footnotesize{$\downarrow$} & N-MOS \footnotesize{$\uparrow$} & SBS \footnotesize{$\uparrow$} & AAE \footnotesize{$\downarrow$} & N-MOS \footnotesize{$\uparrow$} & SBS \footnotesize{$\uparrow$} \\ \midrule
\multicolumn{2}{c}{\ding{55}}  & 4.8 & 4.08 & 0.83 & 5.7 & 4.10 & 0.82 & 5.9 & 4.07 &  0.80  \\ 
\multicolumn{2}{c}{\ding{51}} & 4.8 & \textbf{4.33} & \textbf{0.86} & 5.7 & \textbf{4.30} & \textbf{0.86} & 5.9 & \textbf{4.20} & \textbf{0.84} \\ 
\bottomrule
\end{tabular}
}
\vspace{-6pt}
\end{table}

\begin{table}[t]
    \centering
    \caption{\textbf{Speech denoising results} on LibriMix Test (best score in each column is in bold, second best score is underlined).}
    \label{tab:noise}
    \resizebox{1.\linewidth}{!}{ 
    \begin{tabular}{ccccccc}
    \toprule
          & N-MOS \footnotesize{$\uparrow$} & SIG \footnotesize{$\uparrow$} & BAK \footnotesize{$\uparrow$} & OVRL \footnotesize{$\uparrow$} & COS \footnotesize{$\uparrow$} & ~ \\ \midrule
        \color{gray} Noisy  & \color{gray} 2.62 & \color{gray}3.97 & \color{gray}2.52 & \color{gray} 2.97 & \color{gray} - & ~ \\ 
        DCCRNet~\cite{hu2020dccrn}  & 4.15 & 4.08 & 4.26 & 3.73 & \underline{0.89} & ~ \\
        Conv-TasNet~\cite{luo2019conv}  & 4.12 & 4.18 & \underline{4.30} & 3.78  & \textbf{0.91} & ~  \\
        AnCoGen~\cite{sadok2025ancogen} & 4.24 & 4.21 & 4.32 & \textbf{3.81} & 0.73 & ~ \\
        \rowcolor{blue!10}
        RT-MAE (Ours) & \textbf{4.25} & \textbf{4.23} & 4.29 & \underline{3.80} & 0.86 & ~ \\
        \bottomrule
    \end{tabular}
    \vspace{-6pt}
    }
    
\end{table}

\section{Conclusion}
In this work, we introduced RT-MAE, a residual-token masked autoencoder for speech modeling. By capturing information not represented by explicit attributes, the residual tokens improve reconstruction and enhance robustness to noise. We further regularize these tokens to control information flow, ensuring a balanced contribution with structured attributes. This design supports practical applications such as speech enhancement. 
As future work, a particularly promising application is the use of residual tokens in expressive text-to-speech to capture subtle prosodic and emotional cues for more natural synthesis, while the framework also enables fine-grained voice conversion and style transfer by controlling residual factors alongside explicit attributes.
\\

\noindent\textbf{Acknowledgments:} This work was performed using HPC resources from GENCI–IDRIS (Grant 2025-A0181016041).

\makeatletter
\def\thebibliography#1{%
  \section{REFERENCES}%
  \small
  \linespread{0.90}\selectfont 
  \list{[\arabic{enumi}]}{%
    \settowidth\labelwidth{[#1]}%
    \leftmargin\labelwidth
    \advance\leftmargin\labelsep
    \usecounter{enumi}
    \itemsep 0pt 
    \parsep 0pt
    \topsep 0pt
  }%
  \def\newblock{\hskip .11em plus .33em minus .07em}%
  \sloppy\clubpenalty4000\widowpenalty4000
  \sfcode`\.=1000\relax
}
\def\endthebibliography{%
  \endlist
}
\makeatother

\bibliographystyle{IEEEtran}
\bibliography{refs}

\end{document}